\documentclass[aps,pra,preprint,groupedaddress,showpacs,showkeys]{revtex4}
\usepackage{graphicx}
\usepackage[figuresleft]{rotating}
\usepackage{makecell,rotating}
\usepackage{longtable}

\begin{document}


\title{Gamma -ray spectra in the positron-annihilation process of molecules at room temperature}

\author{Lin Tang}
\author{Xiaoguang Ma}
\email{hsiaoguangma@ldu.edu.cn}
\author{Jipeng Sui}
\author{Meishan Wang}
\author{Chuanlu Yang}
\affiliation{School of Physics and Optoelectronic Engineering, Ludong University, Yantai, Shandong 264025, People's Republic of China}


\date{\today}

\begin{abstract}
In this study, a fully self-consistent method was developed to obtain the wave functions of the positron and electrons in molecules simultaneously. The wave function of a positron at room temperature , with a characteristic energy of approximately $0.04 eV$, was used to analyse the experimental results of its annihilation in helium, neon, hydrogen, and methane molecules.  The interactions between the positron and molecule provide a significant correction in the gamma-ray spectra of the annihilating electron--positron pairs. It was also observed that high-order correlations offered almost no correction in the spectra, as the interaction between the low-energy positron and electrons cannot drive the electrons into excited electronic states. More accurate studies, which consider the coupling of the positron--electron pair states and vibration states of nuclei, must be undertaken.
\end{abstract}


\maketitle

\section{Introduction\label{}}
Gamma-ray spectra in low-energy positron annihilation, usually at room temperature, were measured extensively in many gas-phase molecules earlier\cite{1,2}. In recent years, significant theoretical studies have also improved our understanding of the positron--electron annihilation process in atoms and small molecules\cite{3,4,5,6,7}. However, the theoretically predicted total profiles of gamma-rays for most of the molecules were only within approximately 70\% agreement with the experimental results\cite{2}. The annihilation spectra obtained in the recent low-energy plane-wave positron (LEPWP) approximation \cite{4,5} were always broader than those measured experimentally\cite{1}.

The broadening of the gamma-ray spectra is mostly related to the momentum distribution of the positron--electron pairs in the small momentum region, i.e. the electrons distributed in the lowest momentum region and the low-energy positron play an important role in the annihilation process. However, the understanding of low-energy positrons and their behaviour in molecules is still incomplete as compared to the more familiar subject of electrons.  One of the reasons is that an accurate positron wave function in the annihilation process in molecular systems is difficult to obtain. In the present study, a fully self-consistent method was developed to obtain the wave functions of the positron and electrons in molecules simultaneously. This low-energy positron wave function can account for part of the positron--electron correlations. The corresponding gamma-ray spectra agree very well with the recent experimental results.

\section{The wave function and annihilation of positron--electron pairs\label{}}
Generally, there are four different orbitals: $\alpha^-$, $\beta^-$ electrons and $\alpha^+$, $\beta^+$ positrons in positron-molecule systems under the unrestricted Hartree-Fock (UHF) self-consistent field approximation\cite{8,9}. The Fock operators $\bf{F = H + J - K}$ for electrons and positrons are defined by the single-electron Hamiltonian
\begin{equation}
H_i = \int{\phi^*_i(1)H\phi_i(1)d\vec{r}_1},
\end{equation}
Coulomb integrals
\begin{equation}
J_{ij} = \int\int{\phi^*_i(1)\phi^*_j(2)\frac{1}{r_{12}}\phi_i(1)\phi_j(2)d\vec{r}_1d\vec{r}_2},
\end{equation}
and exchange integrals
\begin{equation}
K_{ij} = \int\int{\phi^*_i(1)\phi^*_j(2)\frac{1}{r_{12}}\phi_j(1)\phi_i(2)d\vec{r}_1d\vec{r}_2},
\end{equation}
between the $i$th and $j$th molecular orbitals. In most of the actual implementations, the molecular orbitals are expansions of the atomic orbital basis functions $\phi_i = \sum_r{\chi_rC_{ri}}$, where $\chi_r$ is a set of atomic orbitals and $C_{ri}$s are the weight coefficients calculated by a self-consistent field procedure.

In this form, the Fock matrices for the $\alpha^-$ and $\beta^-$ electrons are $\bf{F^{\alpha^-} = H^- + J^- - K^{\alpha^-}}$ and $\bf{F^{\beta^-} = H^- + J^- - K^{\beta^-}}$, respectively. The elements between the atomic orbital basis functions $s$ and $r$ for electrons are
\begin{equation}
\langle{\bf{H^-}}\rangle_{rs} = \int{\chi^*_r(1)\{-\frac{1}{2}\nabla^2_1-\sum_{N}\frac{Z_N}{|r_1-R_N|}\}\chi_s(1)d\vec{r}_1}
\end{equation}
\begin{equation}
\langle{\bf{J^-}}\rangle_{rs} = \sum_{tu}(p^{\alpha^-}_{tu}+p^{\beta^-}_{tu}-p^{\alpha^+}_{tu}-p^{\beta^+}_{tu})\int\int{\chi^*_r(1)\chi^*_t(2)\frac{1}{r_{12}}\chi_s(1)\chi_u(2)d\vec{r}_1d\vec{r}_2}
\end{equation}
\begin{equation}
\langle{\bf{K^{\alpha^-}}}\rangle_{rs} = \sum_{tu}p^{\alpha^-}_{tu}\int\int{\chi^*_r(1)\chi^*_t(2)\frac{1}{r_{12}}\chi_u(1)\chi_s(2)d\vec{r}_1d\vec{r}_2}
\end{equation}
for the $\alpha^-$ electrons, and
\begin{equation}
\langle{\bf{K^{\beta^-}}}\rangle_{rs} = \sum_{tu}p^{\beta^-}_{tu}\int\int{\chi^*_r(1)\chi^*_t(2)\frac{1}{r_{12}}\chi_u(1)\chi_s(2)d\vec{r}_1d\vec{r}_2}
\end{equation}
for the $\beta^-$ electrons, respectively. The Coulomb interaction between electrons and positrons is considered in Eq.(5).

The elements of the Fock matrices $\bf{F^{\alpha^+} = H^+ + J^+ -K^{\alpha^+}}$ and $\bf{F^{\beta^+} = H^+ + J^+ - K^{\beta^+}}$ for the $\alpha^+$ and $\beta^+$ positrons are easily found between the positron basis functions as given below:
\begin{equation}
\langle{\bf{H^+}}\rangle_{rs} = \int{\chi^*_r(1)\{-\frac{1}{2}\nabla^2_1+\sum_{N}\frac{Z_N}{|r_1-R_N|}\}\chi_s(1)d\vec{r}_1}
\end{equation}
\begin{equation}
\langle{\bf{J^+}}\rangle_{rs} = \sum_{tu}(p^{\alpha^+}_{tu}+p^{\beta^+}_{tu}-p^{\alpha^-}_{tu}-p^{\beta^-}_{tu})\int\int{\chi^*_r(1)\chi^*_t(2)\frac{1}{r_{12}}\chi_s(1)\chi_u(2)d\vec{r}_1d\vec{r}_2}
\end{equation}
\begin{equation}
\langle{\bf{K^{\alpha^+}}}\rangle_{rs} = \sum_{tu}p^{\alpha^+}_{tu}\int\int{\chi^*_r(1)\chi^*_t(2)\frac{1}{r_{12}}\chi_u(1)\chi_s(2)d\vec{r}_1d\vec{r}_2}
\end{equation}
for the $\alpha^+$ positrons, and
\begin{equation}
\langle{\bf{K^{\beta^+}}}\rangle_{rs} = \sum_{tu}p^{\beta^+}_{tu}\int\int{\chi^*_r(1)\chi^*_t(2)\frac{1}{r_{12}}\chi_u(1)\chi_s(2)d\vec{r}_1d\vec{r}_2}
\end{equation}
for the $\beta^+$ positrons, respectively. Eq.(9) considers the correlation of the electrons with positrons.

The density matrices for the electrons and positrons are given by an iterative calculation of the weight coefficients i.e. $p_{tu}^{\alpha^-} = \sum\limits_{i = 1}^{n_{\alpha^-}}C^*_{ti}C_{ui}$, $p_{tu}^{\beta^-} = \sum\limits_{i = 1}^{n_{\beta^-}}C^*_{ti}C_{ui}$, $p_{tu}^{\alpha^+} = \sum\limits_{i = 1}^{n_{\alpha^+}}C^*_{ti}C_{ui}$, and $p_{tu}^{\beta^+} = \sum\limits_{i = 1}^{n_{\beta^+}}C^*_{ti}C_{ui}$.  Thus, a set of molecular orbitals for electrons and positrons is obtained by the solution of Roothaan equations\cite{8}. The total $\alpha^-$ or $\beta^-$-electron wave function $\Phi^{\alpha^-}$ or $\Phi^{\beta^-}$, respectively, satisfying the Pauli principle, is built up as an anti-symmetrised product of all the above molecular electronic orbitals\cite{9}. In the same way, the total $\alpha^+$ or $\beta^+$-positron wave function $\Phi^{\alpha^+}$ or $\Phi^{\beta^+}$, respectively, obeying the Pauli principle, is expanded in terms of an anti-symmetrised product of one-positron orbitals. As a result, the positron--electron pair orbital is written as a product i.e. $\Phi = \Phi^{\alpha^-}\Phi^{\beta^-}\Phi^{\alpha^+}\Phi^{\beta^+}$. The Slater-determinant ensures that the motion of electrons or positrons with parallel spin is correlated. Moreover, the distinguishability  of the electrons and positrons is represented in the Hartree product form which signifies the simultaneous probability of finding the electrons and positrons at the same point, i.e. the probability of existence of a positron--electron pair.

In the annihilation process, the momentum $\vec{p}$ of the emitting photons is equal to the momentum of the positron--electron pair, i.e. $\vec{p} = \vec{k^-} + \vec{k^+}$. Then, the momentum distribution of photons is obtained by a Fourier transform from the positron--electron pair wave function
\begin{equation}
A(\vec{p}) = \int{\Phi^-({\vec{r}})\Phi^+(\vec{r})e^{-i\vec{p}\cdot\vec{r}}d\vec{r}}
\end{equation}
Alternatively, according to the convolution theorem, the momentum distribution of photons is written as
\begin{equation}
A(\vec{p}) = \phi^-(\vec{k^-})\otimes\phi^+(\vec{k^+})=\int^{+\infty}_{-\infty}{\phi^+(\tau)\phi^-(\vec{p}-\tau)d\tau}
\end{equation}
where,
\begin{equation}
\phi^-(\vec{k^-}) = \int{\Phi^-({\vec{r}})e^{-i\vec{k^-}\cdot\vec{r}}d\vec{r}}
\end{equation}
and
\begin{equation}
\phi^+(\vec{k^+}) = \int{\Phi^+({\vec{r}})e^{-i\vec{k^+}\cdot\vec{r}}d\vec{r}}
\end{equation}
are the wave functions of the electrons and positrons in momentum space, respectively.

The momentum of the electron-positron pair is rotationally averaged in the gas or liquid experiments. Hence, the theoretical momentum distribution must be spherically averaged\cite{1} to enable comparison with the experimental measurements. The radial distribution function in momentum space is defined by
\begin{equation}
D(p) = \int_0^\pi{d\theta}\int_0^{2\pi}{d\phi{P^2}\sin{\theta}|A(\vec{p})|^2}
\end{equation}
where, $P,\theta,\phi$ are the spherical coordinates \cite{7}. Hence, the theoretical spherically averaged momentum distribution is given by
\begin{equation}
\sigma(p) = \frac{D(p)}{4\pi{p}^2}
\end{equation}
which signifies the averaged probability to encounter the electron-positron pair on the surface with momentum $|P|$.

The gamma-ray spectra in the annihilation process undergo a Doppler shift in energy owing to the longitudinal momentum component of the positron--electron pair\cite{7}. Hence, integration over a plane perpendicular to $p$ must be performed to obtain the total probability density at the momentum $p = 2\epsilon/c$. Then, the gamma-ray spectra for the positron--electron pair are given by
\begin{equation}
\Omega(\epsilon) = \frac{1}{c}\int_{2\epsilon/c}^{\infty}{\sigma(p)pdp}.
\end{equation}
The Doppler shift from the centre ($mc^2 = 511$ keV) is given by $\epsilon$.

\section{Application and discussion\label{}}
\begin{figure}
\centering
\includegraphics[width = 1.0\textwidth, angle = 0]{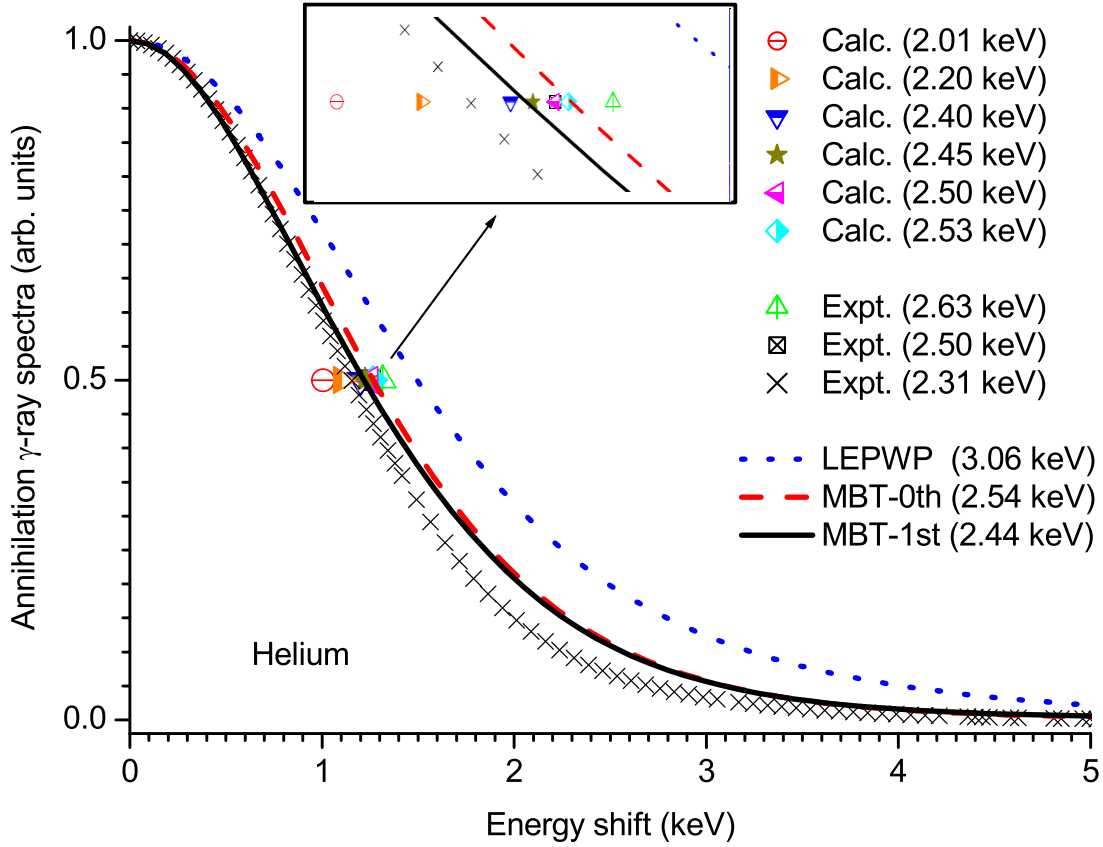}
\caption{\label{HeFig} Gamma-ray spectra in the positron--electron pair annihilation process in helium compared with experimental and theoretical values.}
\end{figure}
Fig. 1 shows a comparison of the present theoretical gamma-ray spectra of helium in the positron--electron annihilation process with other calculated and measured values. A width of 2.63 keV was obtained by measurement of the annihilation radiation \cite{10} and the widths 2.50 keV and 2.31 keV were fitted by one and two Gaussian functions, respectively, for the same measurements\cite{1}. In recent studies, the width 2.31 keV was considered to be relatively accurate and is taken as the reference value to be compared with our theoretical value. The LEPWP approximation gives the width of 3.06 keV\cite{4} indicated by the short blue dashed line. This approximation considered only the electron wave function of atoms or molecules and regarded the positron wave function as a plane wave with zero energy. When the energy of the incoming positron tends to zero, the plane wave is negligible. Therefore, Eq.(12) uses only the electron wave function to calculate the photon momentum\cite{1} as given below:
\begin{equation}
A(\vec{p}) = <\vec{p}|\delta|\mu\nu> = \int{\psi_{\mu}({\vec{r}})\psi_{\nu}(\vec{r})e^{-i\vec{p}\cdot\vec{r}}d\vec{r}} = \int{\psi_{\mu}({\vec{r}})e^{-i\vec{p}\cdot\vec{r}}d\vec{r}}
\end{equation}
where, $\psi_{\mu}({\vec{r}})$ and $\psi_{\nu}(\vec{r})$ are the electron and positron wave functions, respectively. According to previous studies, neglecting the wave function of the positron results in an overestimation of the width by 30 percent.

In many cases the positron was considered to be thermalised before annihilation and the momentum of the positron--electron pair was approximately equal to that of the electron. However, when the positron was close to the atom, the attraction between them increased the speed of the positron slightly; the speed of the electron was also modified to some extent as it was attracted to the positron. Therefore, the momentum distribution of the positron--electron pair was not exactly equivalent to the momentum distribution of the electron in the undisturbed atom. Many approximations, such as polarisation orbit approximation (2.45 keV\cite{16}), S-wave phase shift (2.20 keV\cite{12}), and an atmospheric pressure condition (2.01 keV\cite{11}) were considered in previous studies. As the positrons were disturbed, the electrons were attracted and moved closer to the positrons and farther from the nucleus, thereby increasing the probability of positrons pairing with low momentum electrons. The large discrepancy between the LEPWP line and reference value is probably due to the absence of positron wave functions in Eq.(19).

The long red dashed line with the width 2.54 keV is the spectra obtained using the positron wave function determined by considering the zero-order approximation. As mentioned in Eqs.(1-13), we used the $ab$ $initio$ self-consistent method to solve the Schrodinger equations of the positron and electron at the same time and obtained their self-consistent wave functions. These wave functions were used to determine the momentum distribution of the photon pair. The influence of the positron on the wave function of the electron was considered in the self-consistent process and the wave function of the positron was obtained after considering the influence of the atom on it, hence the gamma-ray spectra are more accurate. In comparison with the LEPWP method, the discrepancy in the short blue line reduced by 22 percent.

Figure 2 shows the schematic diagram of the positron and electron density distribution near the helium atom. The positron density distribution was almost perfectly spherical on a large scale, as shown in Fig. 2(a), whereas on a smaller scale (Fig. 2(b)) it was no longer spherically symmetric. The density of the positron increased noticeably at a distance of several Bohr from the centre. The polarising effect of electrons on the positron density distribution is evident. The wave function of the positron condensed in some directions exhibiting some components of a p-wave characteristic due to the influence of electrons. The electron of the helium atom also showed p-wave components due to the interaction with the incoming positrons as shown in Fig. 2(c). The above analysis indicates that it is necessary to consider the variation and polarisation of the positron and electrons in the theoretical calculation which was confirmed by the results of our calculation.

The black solid line in Fig. 1 represents the gamma-ray spectra obtained by considering the excitation processes of positron and electrons induced by the interaction between the positron and helium atom. As mentioned in reference\cite{3},the first-order correction for the interaction with positrons and electrons is as follows\cite{3}:
\begin{equation}
-\sum_{\mu,\nu}\frac{<\vec{p}|\delta|\mu\nu><\mu\nu|V|n\epsilon>}{\epsilon-\epsilon_{\nu}-\epsilon_{\mu}+\epsilon_n}
\end{equation}
where,
\begin{equation}
<\mu\nu|V|n\epsilon> = \int{\psi^*_{\mu}(\vec{r_1})\psi^*_{\nu}(\vec{r_2})\frac{1}{|\vec{r_1}-\vec{r_2}|}\psi_{\mu}(\vec{r_1})\psi_{\nu}(\vec{r_2})}
\end{equation}
However, the correction of the first term was not apparent as it was only 4 percent approximately. This is because the low-energy positrons did not have sufficient energies to drive the electrons to the excited states, and most of the effects were due to polarisation and not excitation . The width as determined in the present study was 2.44 keV. The widths 2.50 keV\cite{13} and 2.53 keV\cite{1} were obtained by calculations using the variational method, which considered the polarisation effect and 2.40 keV\cite{15} and 2.45 keV\cite{16} were obtained by the polarisation orbit approximation as well as consideration of virtual positive ion (Ps) bubbles.  Thus, the present theoretical results agree well with those of previous studies.

\begin{figure}
\centering
\includegraphics[width = 1.0\textwidth, angle = 0]{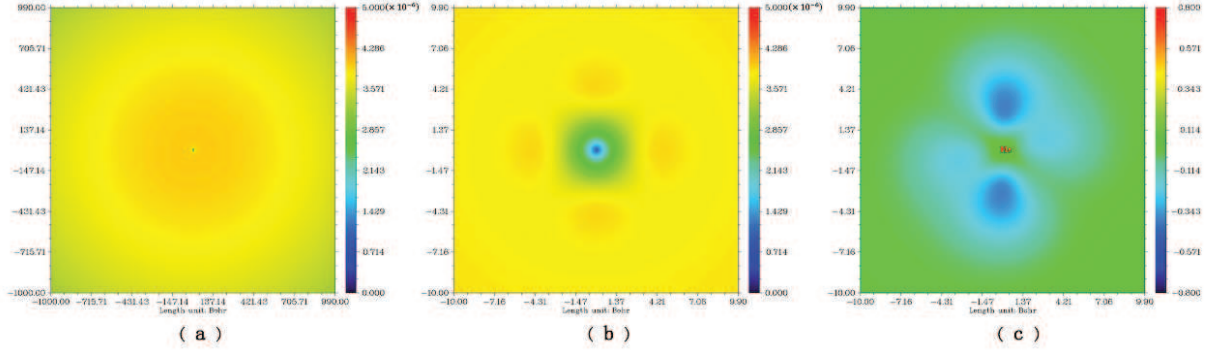}
\caption{\label{NeFig} Positron and electron density distribution near helium atom (The unit of the scale on the graph is Bohr). (a) positron, on a radius scale of 1000 Bohr; (b) positron, on a radius scale of 10 Bohr; (c) electron, on a radius scale of 10 Bohr.}
\end{figure}

Helium has only two electrons that are pulled out easily by the positron hence, the wave function of the positron and correlation effects are of great significance in gamma-ray spectra. As the interaction between the positron and electrons was considered in the calculation of the gamma-ray spectrum, the accuracy of theoretical calculation was vastly improved. From the many-body theory of zero-order and first-order approximations, the widths obtained were 2.54 keV and 2.44 keV, respectively, for helium. The self-consistent electron and positron wave functions were considered by the zero-order approximation, which provided a 17\% correction on the plane-wave approximation. The first-order approximation accounted for the excitation processes of the electrons and positrons, which yielded a width of 2.44 keV with a 20\% correction on the plane-wave approximation. However, the first-order term only had a 4\% correction on the zero-order terms.

\begin{figure}
\centering
\includegraphics[width = 1.0\textwidth, angle = 0]{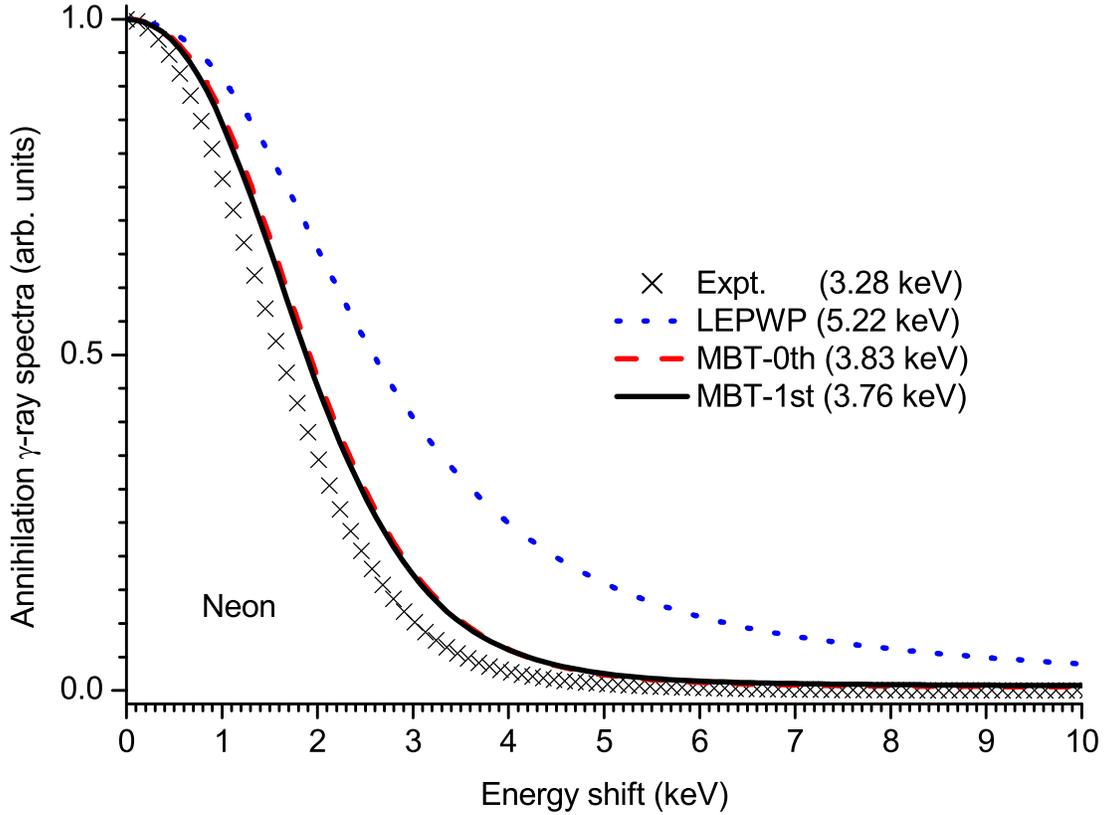}
\caption{\label{NeFig} Theoretical gamma-ray spectra in neon in comparison with other theoretical and experimental values.}
\end{figure}

Fig. 3 shows the theoretical and measured gamma-ray spectra in a neon atom. The width of the experimental gamma-ray spectra fitted by two Gaussian functions and used as reference was 3.28 keV\cite{3}. It is seen that the width of the gamma-spectra produced by the annihilation of electrons and positrons in neon atom are generally larger than that of the helium atom. The value measured by the annihilation radiation was 3.19 keV\cite{9} and the data fitted by one Gaussian function was 3.36 keV\cite{1}. The theoretical value of 2.04 keV was calculated by Doppler widening at an atmosphere\ cite{9}, indicating that the positron wave function was dominant in this annihilation process. A spectral width of 3.28 keV was determined by the variational method\cite{1}. Although 3.32 keV\cite{14} and 3.73 keV\cite{15} were both calculated by using the polarised orbit approximation, 3.32 keV was the width for the solid neon atom. This unusual phenomenon may be due to the larger distribution range of relative momentum of electrons and positrons in the neon atom than that in the helium atom.

\begin{figure}
\centering
\includegraphics[width = 1.0\textwidth, angle=0]{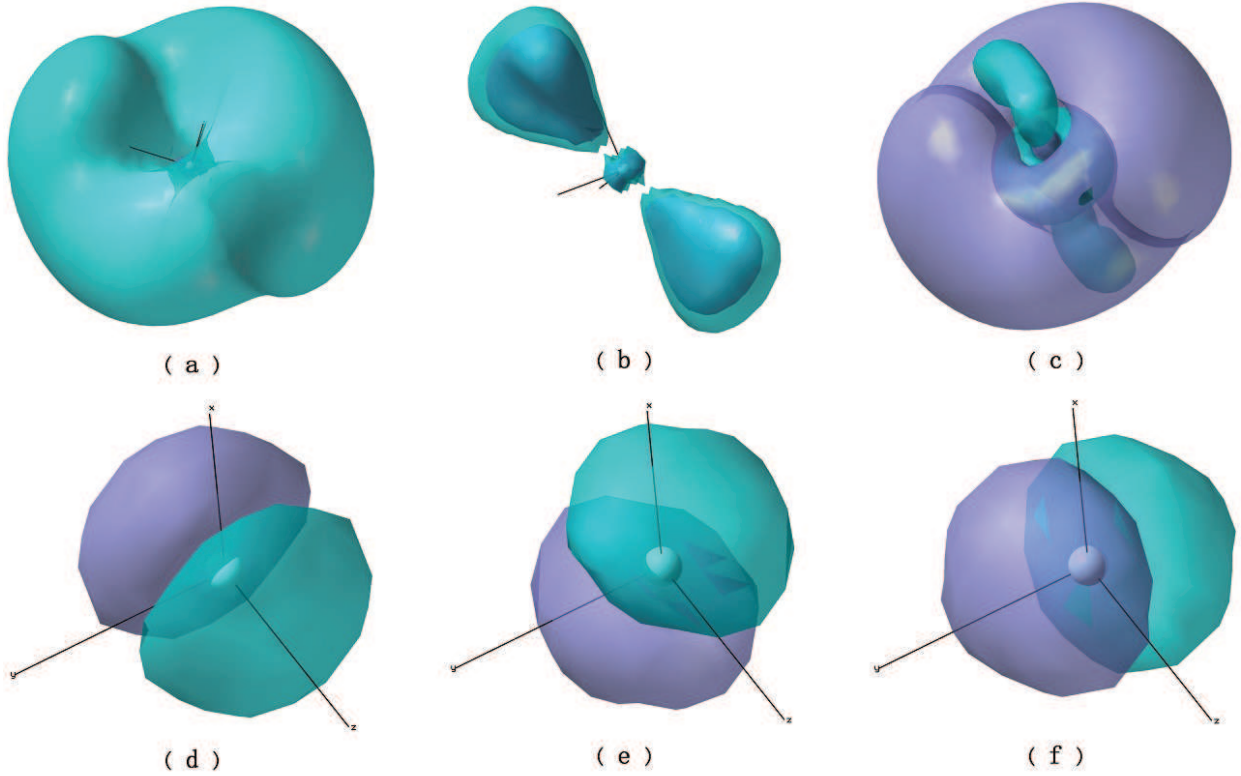}
\caption{\label{H2Fig} Positron and electron density distribution after polarisation, where (a) and (b) represent the density of positrons, (c) represents the density of 2s electrons, (d), (e) and (f) represent the density of 2p$_x$, 2p$_y$, and 2p$_z$ electrons, respectively.}
\end{figure}

As shown in Fig. 4, the wave function of the positron is no longer spherical. This deformation may be due to the influence of the p electron. The wave functions of the 2s electrons under the influence of positrons were polarised to a large extent. The shape of the wave functions of 2p$_x$, 2p$_y$, and 2p$_z$ electrons were basically unchanged; however, their directions were altered after being polarised by the positron, i.e. the p$_x$ wave became a p$_z$ wave, p$_y$ wave became a p$_x$ wave, and p$_z$ wave became a p$_y$ wave. Hence, compared to the results of LEPWP, the errors in the results of the MBT-0th method were reduced by a large percent bringing it close to the reference value. Therefore, it is very important to consider the interaction between the positron and electron in the calculation of gamma-ray spectra. However, the first-order correction was less evident than in the MBT-0th method, hence, the excitation process of the positron and electron may not play an important role in the above annihilation process.

\begin{figure}
\centering
\includegraphics[width = 1.0\textwidth, angle = 0]{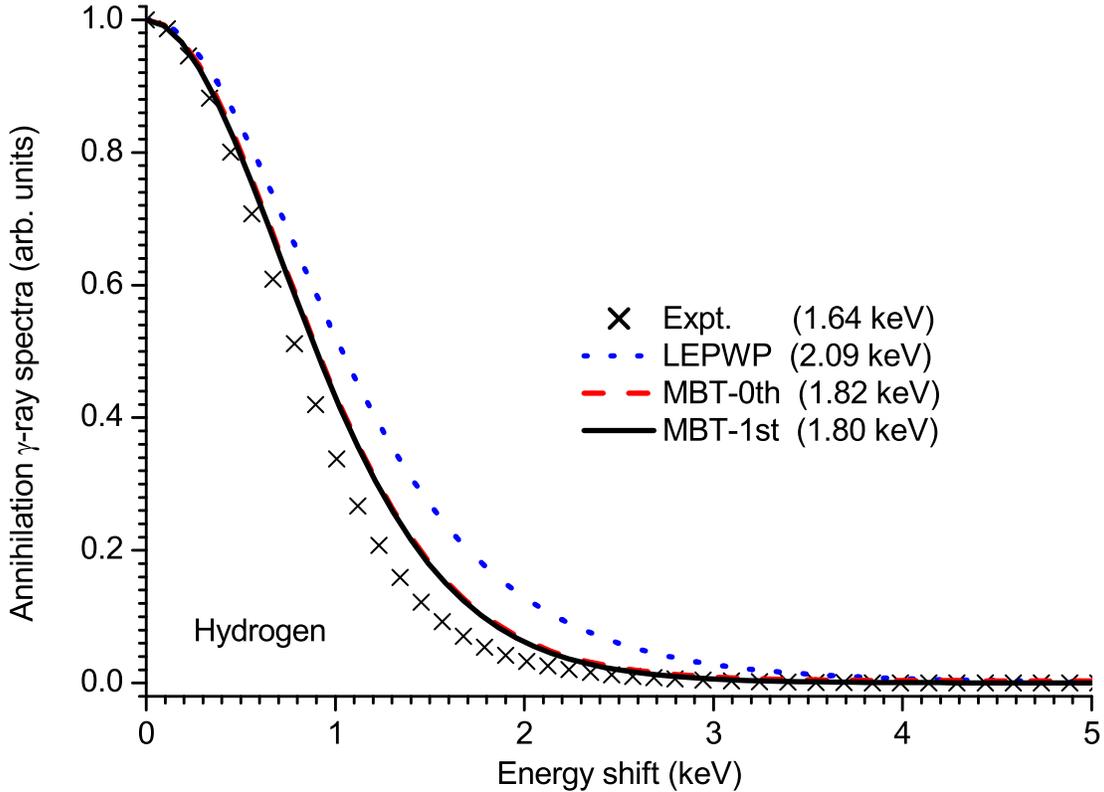}
\caption{\label{H2Fig} Gamma-ray spectra in hydrogen molecule compared with the experimental and other theoretical values. }
\end{figure}

Fig. 5 shows the gamma-ray spectra of positron--electron annihilation in the hydrogen molecule. The experimental value, which was fitted by two Gaussian functions as reference, to compare with the theoretical value, was 1.64 keV\cite{1}. The theoretical calculated gamma-ray spectrum is shown by the three curves in the figure. The short blue dashed line represents the calculation by the LEPWP method, with a width of 2.09 keV\cite{4} and error of 27\%. The spectrum calculated by the MBT-0th method is shown as the long red dashed line with a width of 1.82 keV and an error of 11\%. In comparison with the former two cases of helium and neon, the correction effect of the zeroth order approximation of the positron--electron interaction for the hydrogen molecule was not particularly evident. The spectral line calculated by the MBT-1st method is represented by the solid line in the graph, with a width of 1.80 keV, which was the same as the spectral line in the zero order approximation, and the error was reduced only by 1\%. Therefore, the excitation process of positron--electron annihilation can be considered to be negligible. Comparatively, the data measured by a two-dimensional angular association (2D-ARCR) of annihilation radiation was 1.56 keV\cite{17}. A width of 1.66 keV was measured by the annihilation radiation (ARCR) for liquid H2\cite{18} and 1.71 keV was the width by one Gaussian function fitting\cite{1}. The theoretical width of 1.93 keV was calculated by the static HF method and did not include positron correlation\cite{19}. The width obtained by the polarisation orbit approximation for liquid H2\cite{20} was 1.70 keV. Hence, the coupling of the positron--electron pair and vibration must be considered in future studies.

\begin{figure}
\centering
\includegraphics[width = 1.0\textwidth, angle = 0]{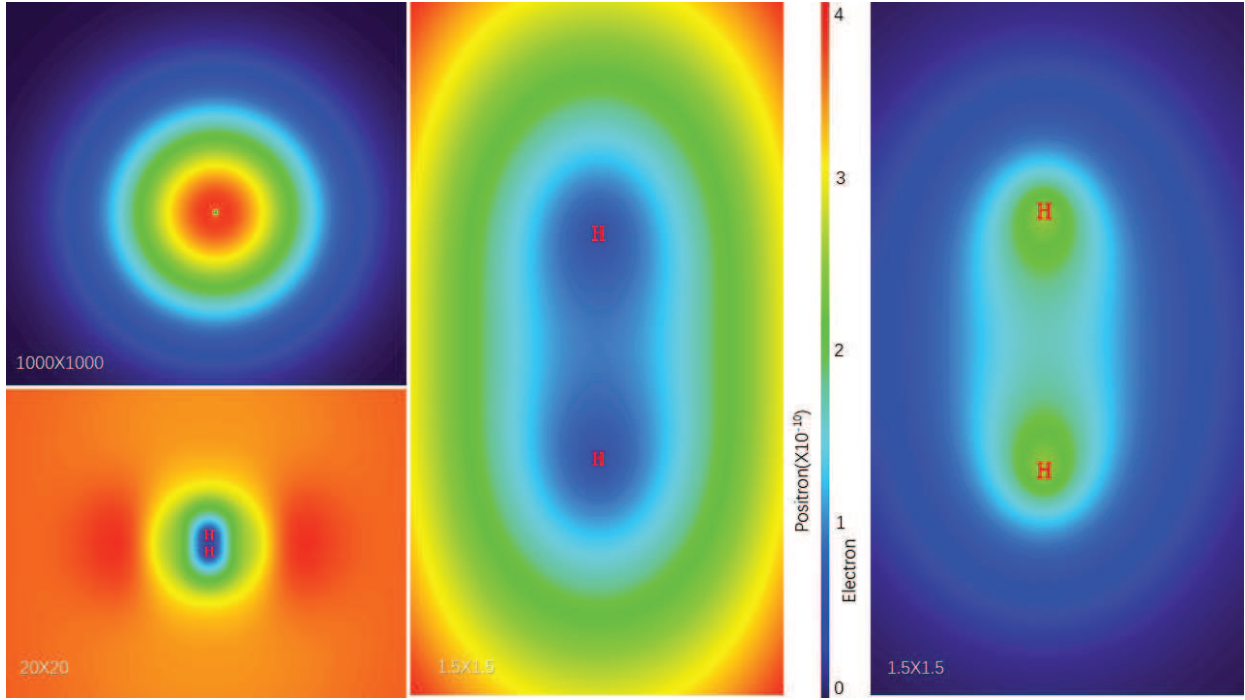}
\caption{\label{H2Fig} Positron and electron density distribution in hydrogen.}
\end{figure}

\begin{figure}
\centering
\includegraphics[width = 1.0\textwidth, angle = 0]{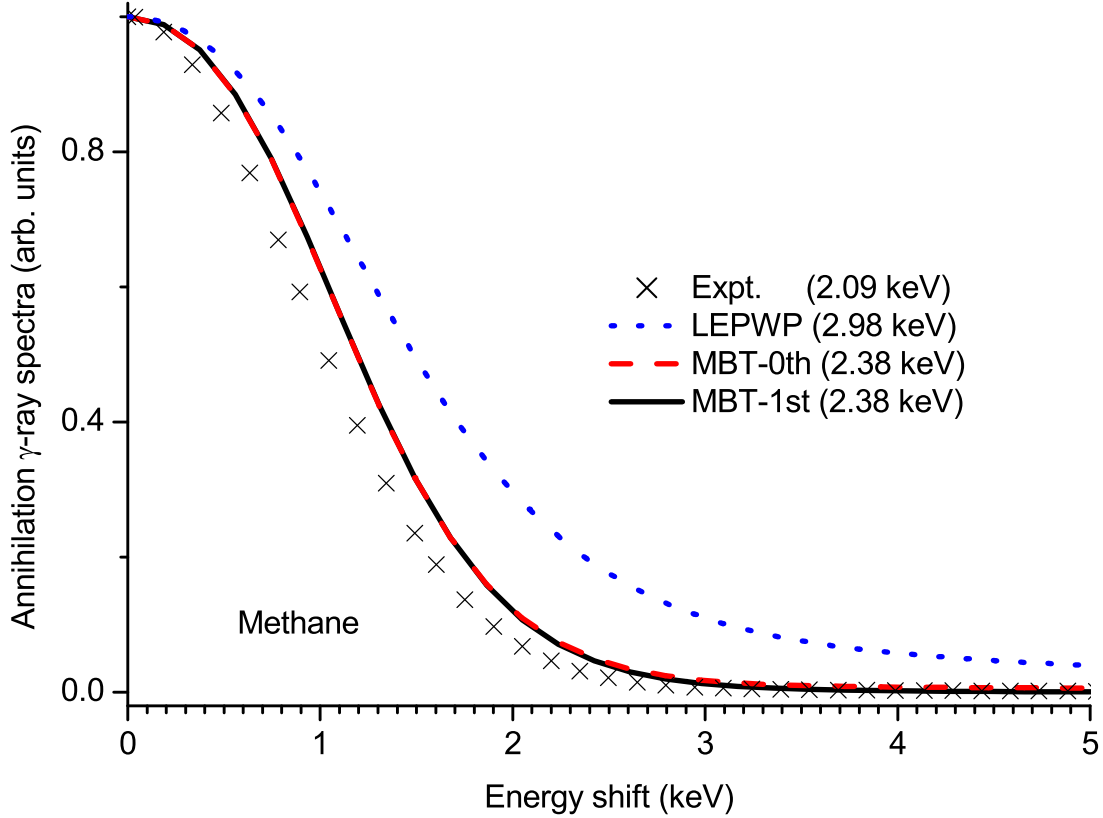}
\caption{\label{CH4Fig} Comparison of theoretical gamma-ray spectrum of methane with experimental results.}
\end{figure}

Fig. 6 shows the schematic diagram of the electron wave functions in positron and hydrogen molecule. The wave function of the positron near the hydrogen molecule is shown on a large scale. It is a perfect spherical wave, except at the origin where it is nearly zero. It is seen that the wave function of the positron is clearly polarised near the hydrogen molecule and its density on the side of the hydrogen-hydrogen bond is higher than that on other parts. The density near the hydrogen atom is the lowest and slightly higher between two hydrogen atoms.

For methane, shown in Fig. 7, the reference value of 1.64 kev\cite{1} was the most recent measurement. The theoretical values calculated by LEPWP, MBT-0th, and MBT-1st methods were 2.98 keV\cite{4}, 2.38 keV, and 2.38 keV, with errors of 43\%, 14\%, and 14\%, respectively. Therefore, compared to the LEPWP method without considering the positron effect, there was a significant improvement in the accuracy of the results of the MBT-0th method. However, as the positrons did not have sufficient energy to excite the electrons in the methane molecules, the MBT-1st method showed no evident improvement in the calculation as compared to the MBT-0th method. Fig. 8 shows the positron and electron density distributions in methane. On a larger scale, the positron density distribution is a perfect spherical wave. The details are seen on a smaller scale, where the positron is seen to be evidently polarised. The first-order correction had little effect on the final results as the excitation of positrons and electrons did not occur in the annihilation process.

\begin{figure}
\centering
\includegraphics[width = 1.0\textwidth, angle = 0]{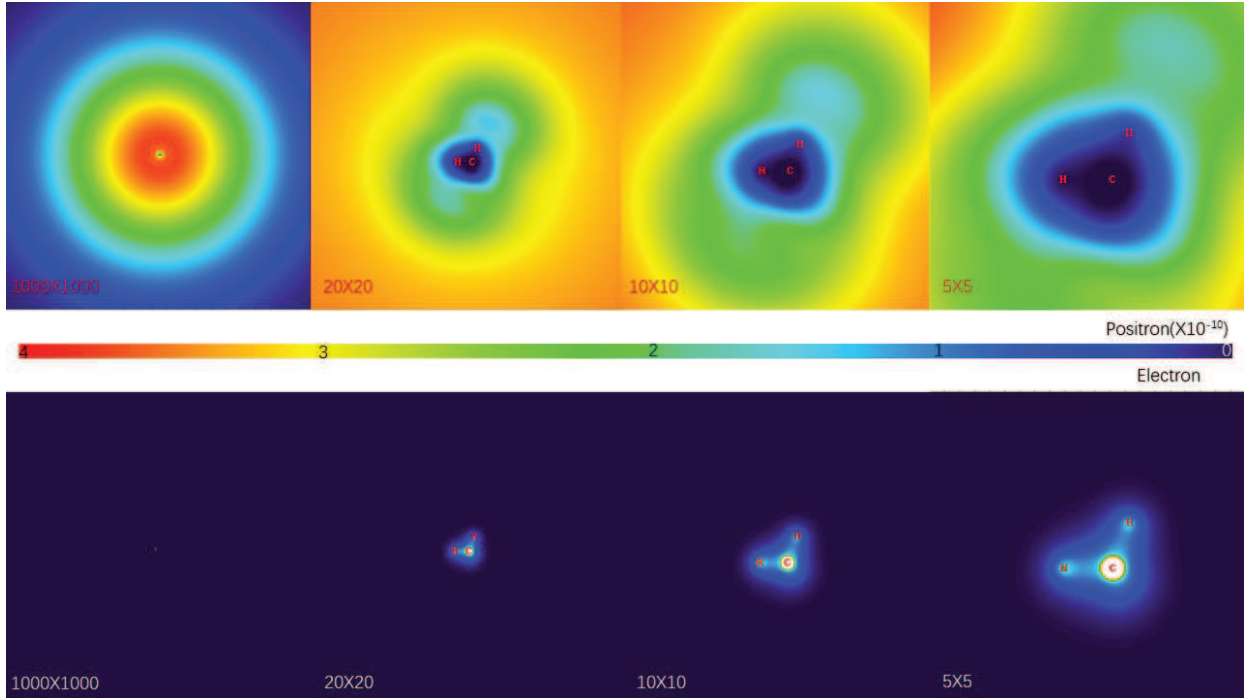}
\caption{\label{CH4Fig} Positron and electron density distribution in methane.}
\end{figure}

\section{Conclusions\label{}}
The present work is the first step to obtain an accurate wave function of the positron in the annihilation process of many-atomic molecules. The fully self-consistent method was developed to study the positron annihilation process in molecules. The results showed that the present method incorporates the wave functions of the electrons and positron appropriately. The interaction and polarisation between the incoming positron and electrons were considered precisely in this method. In the present scheme, the one-body term included the correlation potential which accounted for the polarisation and orientation effects in the positron--electron annihilation process. The wave function of the positron determined by the present scheme provided a significant correction in the gamma-ray spectra (nearly 30 percent). The present theoretical method showed that an accurate wave function of the positron is very important to explain the gamma-ray spectra in molecules. As the positron in the experiments was cooled below the formation threshold of positronium, the excitation processes for higher-order interactions had very little effect and this was proved in the present work.

\section{Acknowledgement\label{}}This work was supported by the National Natural Science Foundation of China under grants No. 11674145 and Taishan Scholars Project of Shandong province (Project No. ts2015110055).


\end{document}